\documentclass[sigconf]{acmart}

\AtBeginDocument{%
  }

\copyrightyear{2027}
\acmYear{2027}
\setcopyright{cc}
\setcctype{by}
\acmConference[WSDM '27]{Proceedings of the Twentieth ACM International Conference on Web Search and Data Mining}{February 15--19, 2027}{Hong Kong, Hong Kong}
\acmBooktitle{Proceedings of the Twentieth ACM International Conference on Web Search and Data Mining (WSDM '27), February 15--19, 2027, Hong Kong, Hong Kong}
\acmDOI{10.1145/3862322.3863856}
\acmISBN{979-8-4007-3117-4/2027/02}

\usepackage{colortbl}
\definecolor{headteal}{RGB}{220,239,235}
\definecolor{rowgray}{RGB}{245,245,245}

\begin{document}

\title{BEACON: Behavior-Anchored Cross-Source Knowledge Graph Construction for Cyber Threat Intelligence}

\author{Changze Li}
\email{changzeli@vt.edu}
\affiliation{%
  \institution{Virginia Tech}
  \city{Blacksburg}
  \country{United States}
}
\author{Yutong Cheng}
\email{yutongcheng@vt.edu}
\affiliation{%
  \institution{Virginia Tech}
  \city{Blacksburg}
  \country{United States}
}
\author{Tsania Camila Finnisa}
\email{111202214241@mhs.dinus.ac.id}
\affiliation{%
  \institution{Dian Nuswantoro University}
  \city{Semarang}
  \country{Indonesia}
}
\author{Qian Cui}
\email{cuiqia@amazon.com}
\affiliation{%
  \institution{Amazon}
  \city{Seattle}
  \country{United States}
}
\author{Wei Ding}
\email{dingwe@amazon.com}
\affiliation{%
  \institution{Amazon}
  \city{Seattle}
  \country{United States}
}
\author{Peng Gao}
\email{penggao@vt.edu}
\affiliation{%
  \institution{Virginia Tech}
  \city{Blacksburg}
  \country{United States}
}
\renewcommand{\shortauthors}{Changze Li et al.}

\begin{abstract}

Cyber threat intelligence (CTI) is foundational to modern cyber defense, yet much of it resides in unstructured reports whose volume and heterogeneity far exceed manual analysis, motivating research on automatically constructing knowledge graphs from CTI reports. However, most existing approaches extract partial information within a single report, leaving the cross-source setting unexplored, where the same threat is given unrelated names. 
Our key insight is that attack behaviors, once mapped to MITRE ATT\&CK (a standardized catalog of attack techniques maintained by experts), can anchor the rest of a report. Attack behaviors are the adversarial actions a report describes, while contextual entities (e.g., threat actors, campaigns, and affected products) and Indicators of Compromise (IoCs; e.g., IP addresses) are their participants and traces. Attaching them to these anchors places every per-report graph in one canonical space, where overlapping techniques become the signal that aligns unrelated names.

We realize this insight in BEACON, an LLM-driven framework for cross-source CTI knowledge graph construction. Its first stage extracts each report into a graph under a \emph{propose-then-verify paradigm}, grounding candidates in report evidence and official ATT\&CK definitions, to suppress LLM misclassification and hallucination. Its second stage merges these graphs with a \emph{hierarchical alignment strategy} that applies signals in decreasing order of determinism, from character-level and semantic similarity to overlapping technique neighborhoods, \emph{iterating} as merges pool neighborhoods. 
No existing benchmark links entities to technique anchors or provides cross-source alignment ground truth. We therefore construct and release two human-annotated datasets from 34 sources: to our knowledge the largest for report-level CTI extraction (8,395 elements) and the first for cross-source consolidation (3,487). On them, BEACON outperforms all baselines by at least 23\% and 9\%, respectively.

\end{abstract}



\maketitle

\section{Introduction}
\label{sec:introduction}

Cyber threat intelligence (CTI) is evidence-based knowledge about existing and emerging cyber threats~\citep{mavroeidis2017cyber}, and it is foundational to modern cyber defense, shifting organizations from reactive incident response to proactive threat anticipation. Much of this knowledge resides in natural-language reports that security vendors publish independently and continuously~\citep{orbinato2022automatic,johnson2016guide,li2022attackg,suarez2026cti}. Reports mainly describe three kinds of information: (1) \emph{Contextual entities} represent high-level context, such as threat actors, campaigns, malware families, and affected products. (2) \emph{Attack behaviors} are mid-level tactics, techniques, and procedures (TTPs)~\citep{strom2020mitre} performed by the adversary. MITRE ATT\&CK~\citep{strom2020mitre} standardizes attack behaviors into a taxonomy of techniques maintained by security experts. Each technique, or TTP, has a unique ID, such as T1190 for exploiting a public-facing application and T1566 for phishing. (3) \emph{Indicators of Compromise (IoCs)} represent low-level traces of the attack procedures, such as IP addresses, file hashes, and CVE identifiers.

These reports differ in wording, naming conventions, and level of detail, and their volume and heterogeneity far exceed the capacity of manual analysis. This motivates a line of work that organizes this knowledge into graphs for large-scale querying~\citep{hogan2021knowledge,gao2023threatkg,cheng2025ctinexus}, but \textbf{two major limitations remain}. \textbf{First}, existing methods cover only \textbf{part of the information} in a CTI report, and fall into two research focuses. One maps attack behavior descriptions in natural language to ATT\&CK techniques, but leaves out the surrounding contextual entities and IoCs~\citep{satvat2021extractor,li2022attackg,cheng2025crucialg}. The other extracts contextual entities and IoCs as relation triplets in each report's own wording, without mapping behaviors to ATT\&CK techniques or capturing their relations to them~\citep{cheng2025ctinexus,gao2023threatkg,marchiori2023stixnet,yang2026cti}. No existing work extracts both into one structure. Recent work turns to large language models (LLMs) for their stronger natural-language processing capability~\citep{zhang2025attackg+,cheng2025ctinexus,yang2026cti}, but three properties of CTI make extraction difficult. The same entity can play different roles in different contexts: Gmail can be an affected product or the tool used to send phishing emails, and Microsoft can be the affected vendor or the report's own publisher, which should not be extracted. ATT\&CK contains hundreds of techniques whose definitions differ in fine-grained details: lateral movement through remote services is T1021~\citep{mitreT1021} with legitimate access but T1210~\citep{mitreT1210} through a vulnerability. And LLM outputs are generated rather than retrieved from the text, so extracted information may not exist in the source. These properties make LLM-based extraction suffer from misclassification and hallucination~\citep{buchel2025sok,mezzi2025large,tao2024context}.

\begin{figure*}[t]
  \centering
  \includegraphics[width=\textwidth]{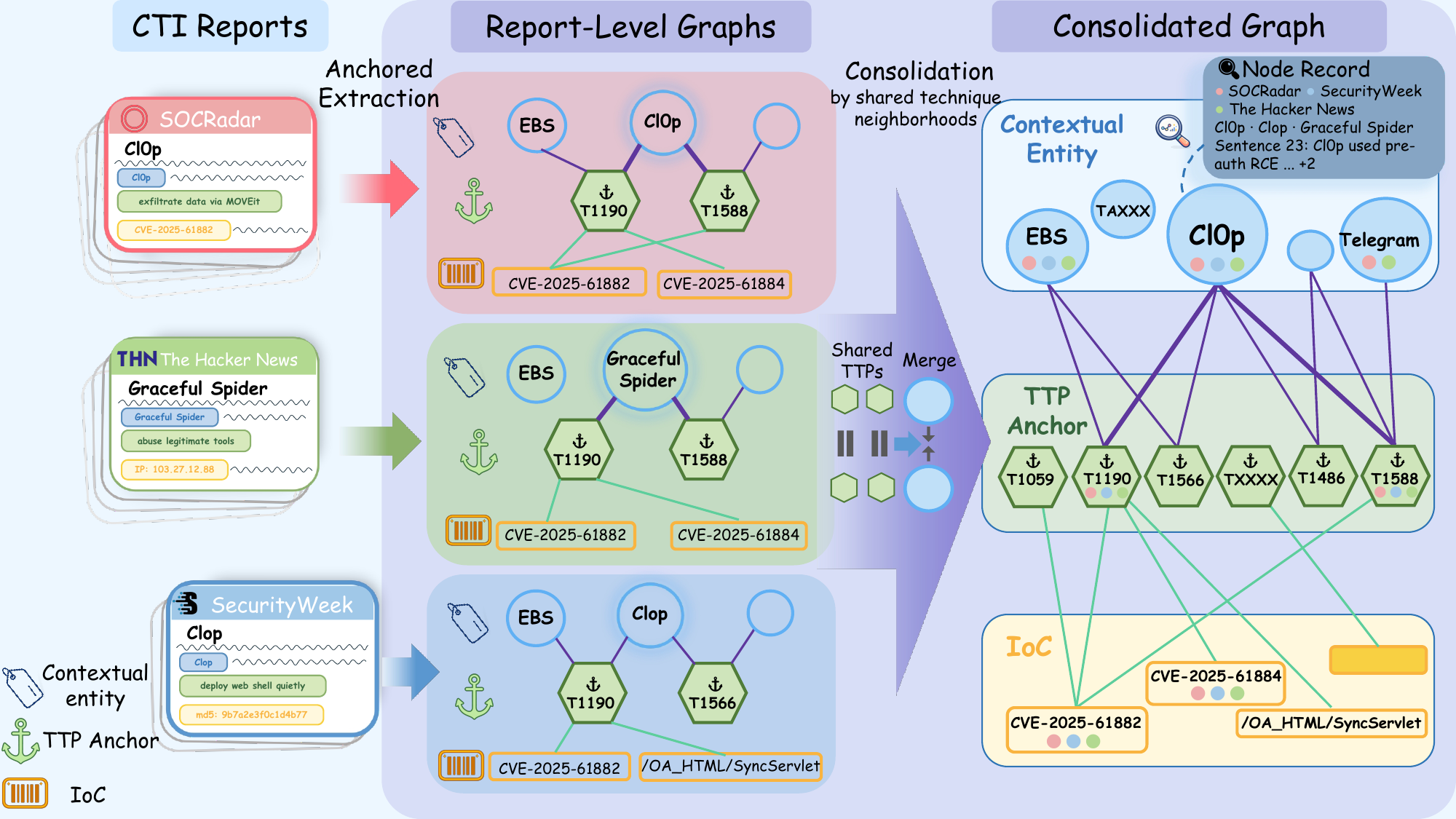}
  \caption{From heterogeneous reports to one consolidated graph. Left: three major CTI sources, SOCRadar~\citep{socradar2026blog}, The Hacker News~\citep{hackernews2026}, and SecurityWeek~\citep{securityweek2026}, cover the same campaign and name its actor \emph{Cl0p}~\citep{ernalbant2025clop}, \emph{Graceful Spider}~\citep{lakshmanan2025clop}, and \emph{Clop}~\citep{kovacs2025clop}. Middle: the space-anchoring stage extracts each report into a graph in which contextual entities (circles) and IoCs (rectangles) attach to ATT\&CK technique anchors (hexagons). Right: the consolidation stage merges nodes that refer to the same entity, where shared technique neighborhoods align the unrelated name. The colored dots inside a node mark its source reports, and the merged actor node keeps every alias, source, and evidence sentence in its record.}
  \label{fig:overview}
  \Description{Three-part diagram read left to right. On the left, three stacked report cards from SOCRadar, The Hacker News, and Security Week describe the same campaign. Highlighted phrases show each vendor's actor name (Cl0p, Graceful Spider, Clop), one behavior sentence, and one indicator such as CVE-2025-61882. Arrows labeled Anchored Extraction lead to the middle, where each report appears as its own small graph: circular contextual-entity nodes such as EBS and the Cl0p, hexagonal ATT\&CK technique anchors such as T1190, T1566, and T1588, and rectangular IoC nodes such as CVE-2025-61882 and a web path. Edges connect entities and IoCs only to anchors. A large arrow labeled Consolidation, with the note shared technique neighborhoods, leads to the right, where the consolidated graph arranges nodes into three horizontal sections labeled Contextual Entity, TTP Anchor, and IoC. Nodes shared across reports, such as Clop, T1190 and CVE-2025-61882, are merged into a single node, with colored dots inside each node indicating which sources reported it. The three actor names have merged into one node, and a magnified node record lists its aliases Cl0p, Clop, and Graceful Spider, its three sources, and a supporting evidence sentence.} 
\end{figure*}

\textbf{Second}, no existing method reliably constructs a knowledge graph \textbf{across} reports from \textbf{different} CTI sources. Vendors independently report on the same threat under different naming conventions, and no single report contains all aliases used across sources~\citep{saha2025expert,cheng2026cticonnect}. For instance, three reports on an Oracle E-Business Suite campaign name the same threat actor \emph{Cl0p}~\citep{ernalbant2025clop}, \emph{Clop}~\citep{kovacs2025clop}, and \emph{Graceful Spider}~\citep{lakshmanan2025clop} (Figure~\ref{fig:overview}, left): the first two differ by a character, while the third is entirely unrelated. Most existing work deduplicates entities within a single report~\citep{cheng2025ctinexus}, and the few multi-report attempts rely on embedding similarity~\citep{yang2026cti}, which cannot align unrelated names such as Cl0p and Graceful Spider. While LLMs can match entities from context, comparing all entity pairs across reports is costly, and LLMs can incorrectly merge distinct entities that appear in related narratives. For example, reports on Oracle E-Business Suite also mention the actor's earlier MOVEit campaign as background~\citep{ernalbant2025clop,lakshmanan2025clop}, and an LLM comparing reports can incorrectly merge the two campaigns since they share the same actor and context.

In this work, we aim to construct one knowledge graph from CTI reports across sources, which covers all three kinds of information and reliably merges nodes that refer to the same real-world entity. Our \textbf{key insight} is that attack behaviors, a central component of CTI reports, can be mapped to ATT\&CK techniques, however a source words them, while contextual entities and IoCs describe the participants and traces of these behaviors~\citep{satvat2021extractor,li2022attackg}. Attaching entities to ATT\&CK techniques as \emph{anchors} therefore captures the relationships that existing methods leave unextracted, and unifies the per-report graphs in a shared canonical space, giving entities with unrelated names an additional domain-specific signal for alignment.

We build on this insight with \textbf{BEACON} (BEhavior-Anchored CONsolidation of CTI reports), an LLM-driven two-stage framework that constructs one knowledge graph across different CTI sources (Figure~\ref{fig:overview}). To suppress misclassification and hallucination, we design a \emph{propose-then-verify paradigm}. The first stage, space-anchoring, extracts each report into a graph in which contextual entities and IoCs attach to ATT\&CK technique anchors. Since attack behaviors mapping to techniques in narratives are interleaved, the LLM first decomposes each report into atomic behaviors grounded in report evidence. It then proposes multiple candidate techniques for each atomic behavior to expand the coverage among hundreds of fine-grained definitions, and verifies each candidate against its official ATT\&CK definition. Under the same propose-then-verify paradigm, contextual entities and IoCs are extracted and attached to the technique anchors when their evidence overlaps.

For the second stage, consolidation, we design a \emph{hierarchical alignment strategy} that merges the per-report graphs by applying signals in decreasing order of determinism, with each level building on earlier merges and every merge verified by the LLM against report evidence. To cover naming divergence from a single character to entirely unrelated names, it matches entities from character-level and semantic similarity to a signal that only the anchored space provides: the overlap of their ATT\&CK technique neighborhoods, the sets of techniques they attach to. Since entities sharing multiple attack behaviors are likely the same, the neighborhood level can match entities with unrelated names, and iterates as each merge pools its members' neighborhoods and exposes new candidates.

No existing benchmark links entities to technique anchors or provides cross-source alignment ground truth. We therefore construct and release two datasets: \textbf{BEACON-Single}, which contains 150 reports from 15 publishers with 8,395 nodes and edges, and to our knowledge is the largest human-annotated benchmark for report-level CTI extraction; and \textbf{BEACON-Group}, which organizes 100 reports from 31 publishers into 33 groups each covering the same or closely related threats, and is the first benchmark for cross-source CTI consolidation. Both datasets are curated through dual expert annotation with senior adjudication. BEACON outperforms all baselines on both tasks, by at least 23\% on extraction and 9\% on consolidation. On the hardest cases, entities that appear under differently spelled or entirely unrelated names across sources, BEACON outperforms all baselines by at least 27\%. Ablation experiments show that ATT\&CK technique neighborhoods are critical for aligning these hard entities. Our code and datasets are available at \url{https://github.com/peng-gao-lab/beacon-cti}.

\section{Related Work}
\subsection{CTI Extraction} 

Evolving CTI standards such as STIX~\citep{oasis2021stix}, MITRE ATT\&CK~\citep{strom2020mitre}, and OpenCTI~\citep{opencti_docs} provide shared vocabularies for representing and exchanging threat knowledge, yet the CTI community often distributes intelligence as natural-language reports. Bridging this gap, one line of work reconstructs attack behaviors from reports and maps them to ATT\&CK techniques. Early methods rely on rule-guided NLP or supervised classifiers
~\citep{husari2017ttpdrill,li2022attackg,legoy2020automated,orbinato2022automatic}, but they struggle with natural-language variability and transfer poorly as threats evolve. Later works leverage pretrained language models such as BERT~\citep{devlin2019bert}, either embedding report text and TTP definitions into a shared space for semantic comparison~\citep{alam2023looking,nguyen2024noise,kumarasinghe2024semantic} or mapping text to TTP IDs end-to-end~\citep{caitexts,li2024automated}. 
They handle variation better but remain bound to pretrained backbones and labeled data. Recent work turns to LLMs~\citep{buchel2025sok}, with TechniqueRAG~\citep{lekssays2025techniquerag} reranking candidate techniques with an instruction-tuned LLM before mapping with a fine-tuned one and AttacKG+~\citep{zhang2025attackg+} expanding triplets into temporal behavior graphs, whose entities carry no relation to the mapped techniques. Haque et al. aggregate predicted TTP IDs into campaign-level sets~\citep{haque2026beyond}, confirming the value of multi-report settings while leaving entities unaligned. Across this line, extraction focuses on attack behaviors, and contextual entities and IoCs are not extracted or not attached to the mapped techniques, while BEACON extracts both and attaches them to these techniques.

Another line of work targets contextual information. These methods extract the named participants of an attack, such as threat actors, malware families, and organizations, and their relations into a graph. Early systems use rule-based and unsupervised NLP techniques~\citep{park2022full}, while recent work extracts ontology-constrained relation triplets with LLMs~\citep{cheng2025ctinexus,zhang2024extract}. Systems such as CTINexus~\citep{cheng2025ctinexus} and CTI-Thinker~\citep{yang2026cti} extract ATT\&CK technique IDs only when the report writes out the ID, such as T1190. LADDER extracts both TTPs and contextual entities, but its ontology only connects each TTP to malware and cannot capture the relations between other entities and attack behaviors~\citep{alam2023looking}. Across this line, extraction focuses on contextual entities and IoCs, without mapping attack behaviors to techniques or capturing their relations with them. Both lines are limited to single reports, while BEACON constructs one graph from reports across sources that covers all three kinds of information.

\subsection{Cross-Source Linking and Alignment} 

Classical entity matching compares names, textual fields, and attributes across sources using similarity functions and binary classifiers~\citep{cohen2002learning,bilenko2003adaptive}, but performs poorly when these fields do not overlap. Neural entity matching encodes records into learned representations for pairwise comparison~\citep{mudgal2018deep,fu2021hierarchical,li2020deep,jiang2024unlocking},
yet textual semantics alone cannot reliably identify the same entity when two records differ significantly. For example, ``Twitter'' and ``X'' refer to the same product, but no embedding can relate the new name to the old one until sufficient training data about the renaming appears. When entities are connected in graphs, structure offers an additional matching signal
~\citep{lacoste2013sigma,suchanek2011paris}. Recent work uses LLMs to judge whether two records refer to the same entity~\citep{wang2025match}, but these judgments still rest on the same textual evidence.

In CTI, a few approaches try to address the entity alignment problem. CTINexus~\citep{cheng2025ctinexus} limits entity alignment to single-report deduplication, and CTI-Thinker~\citep{yang2026cti}, though aligning across sources, relies on embedding similarity. Both fail on CTI naming conventions. Threat entities are often given unrelated codewords, such as Wicked Panda and Brass Typhoon for the same actor, which are distant in both semantics and spelling, so neither embedding nor string similarity can match them. Threat entities can also be given alphanumeric labels, such as APT41 and TA415, which carry no semantics for embeddings to compare and share few characters for string similarity to match. String similarity can further incorrectly merge two labels when their spellings are nearly identical, such as APT41 and APT40. Naming conventions also differ across vendors. For example, Mandiant tracks the actor above as APT41, while Microsoft names it Brass Typhoon in its own weather-themed scheme. Aligning this actor across sources therefore requires matching APT41 with Brass Typhoon, names from two unrelated naming schemes, which neither exact-name nor similarity-based matching can handle. BEACON instead aligns entities by their shared ATT\&CK technique neighborhoods, a signal that is consistent across sources and independent of naming conventions.

\section{Methodology}
\label{sec:methodology}

\begin{figure*}[t]
  \centering
  \includegraphics[width=\textwidth]{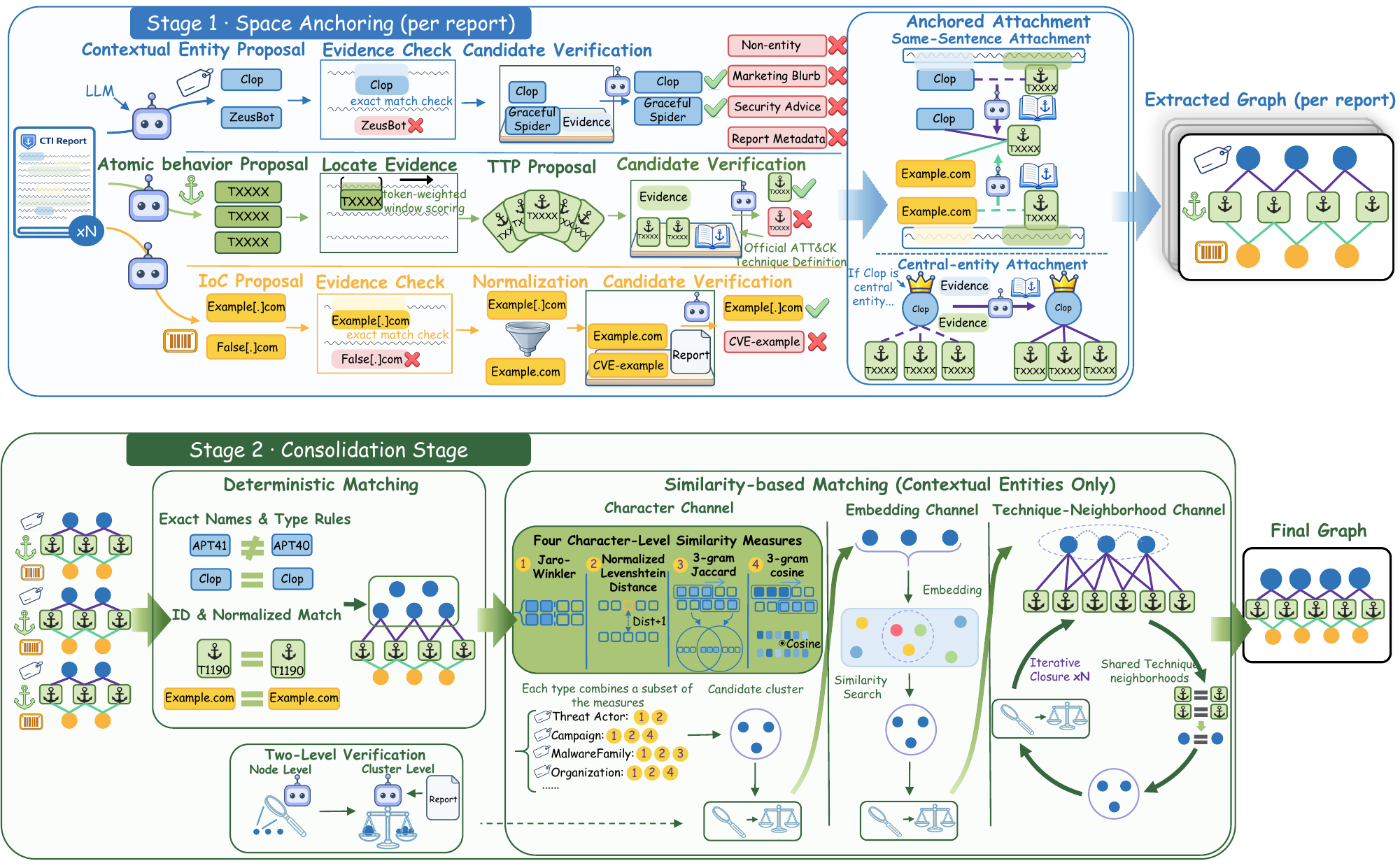}
  \caption{The BEACON pipeline. Stage 1 turns each report into an anchored graph along three tracks: the LLM proposes contextual entities, atomic behaviors, and IoCs with grounded evidence. The candidate techniques of each atomic behavior are verified against official ATT\&CK definitions. IoCs are normalized before verification. Edges are then proposed between verified nodes that share a sentence, each confirmed by the LLM, and report-central entities are additionally paired with every anchor. Stage 2 consolidates the per-report graphs: deterministic matching merges identical technique IDs, normalized IoC forms, and exact names, and the remaining contextual entities enter three candidate channels: character similarity, embedding similarity, and shared technique neighborhoods. Every candidate cluster passes node-level and cluster-level LLM verification, and the neighborhood channel iterates until no new merge is accepted.}
  \label{fig:pipeline}
  \Description{Flow diagram with two large blocks. The upper block, Stage 1 Space Anchoring, shows one CTI report splitting into three horizontal tracks. In the contextual-entity track, LLM-proposed names such as Clop pass an exact-match evidence check, and a verification step removes non-entities, marketing blurbs, security advice, and report metadata. In the behavior track, atomic behaviors are located in the text by token-weighted window scoring, a fan of candidate techniques is proposed, and each candidate is verified against its official ATT\&CK technique definition, drawn as a book icon. In the IoC track, values such as a defanged domain pass an exact-match evidence check, a normalization funnel restores standard forms, and a final verification removes invalid entries. A panel on the right shows anchored attachment: edges proposed from shared sentences are confirmed by the LLM against technique definitions. Below, an entity marked with a crown icon is first judged by the LLM to be the report's central entity, and is then paired with every anchor, with each pairing confirmed by the same LLM verification. The output is a stack of per-report extracted graphs. The lower block, Stage 2 Consolidation, begins with deterministic matching, where identical technique IDs, identical normalized IoCs, and exact names merge, and type rules keep same-scheme names with different numbers apart. Remaining contextual entities pass through three channel panels: a character channel combining a weighted subset of Jaro-Winkler, normalized Levenshtein, 3-gram Jaccard, and 3-gram cosine per entity type; an embedding channel with similarity search; and a technique-neighborhood channel, where an icon of two matching technique anchors beside the arrow indicates that two entities sharing anchors are proposed as candidates, and a dashed outline encloses the entities entering the candidate set. Each channel feeds a two-level LLM verification, drawn as two icons: a scanning machine for the node level, where every member is checked individually, and a balance scale with a report for the cluster level, where the LLM judges whether the whole cluster refers to one entity. The neighborhood channel loops back for iterative closure. The final consolidated graph appears on the right.}

\end{figure*}

Figure~\ref{fig:pipeline} gives an overview of BEACON, an LLM-driven framework that runs in two stages. The first stage, space-anchoring, extracts each of the CTI reports $\{r_1, \dots, r_n\}$ into a graph in which contextual entities and IoCs attach to ATT\&CK technique anchors. The second stage, consolidation, merges nodes that refer to the same real-world entity across reports, producing one consolidated graph $G^{*}$ that covers all three kinds of information. To suppress misclassification and hallucination, both stages follow a propose-then-verify paradigm, where every proposal is verified before entering the graph.

\subsection{Space-Anchoring Stage}
\label{sec:anchoring}

The anchoring stage extracts a CTI report $r$ into a graph $G=(V_a \cup V_c \cup V_o,\, E)$ with three types of nodes: technique anchors $V_a$, mapped to an ATT\&CK technique; contextual entities $V_c$, typed as ThreatActor, Campaign, MalwareFamily, Tool, Organization, Product, Location, or Sector; and IoCs $V_o$, typed as IP addresses, domains, URLs, emails, file paths, Registry keys, file hashes, CVEs, mutexes, or AES keys. The edge design follows two properties of CTI reports. First, attack behaviors are a central component of CTI reports, where contextual entities and IoCs enter the narrative as their participants and traces~\cite{satvat2021extractor,li2022attackg}. Second, attack behaviors can be reliably mapped to MITRE ATT\&CK, however a source words them. Therefore, every edge in $E$ connects a contextual entity or IoC to a technique anchor, which gives the consolidation stage a domain-specific signal for cross-source alignment. These node types and edges form the predefined ontology of BEACON.

\subsubsection{Behavior Grounding}
BEACON maps the attack behaviors in report $r$ to ATT\&CK techniques in four steps. First, since CTI reports describe attacks as continuous narratives, where actions that map to different techniques appear without clear boundaries, the LLM decomposes the report into a list of \emph{atomic behaviors}. We define an atomic behavior $b$ as one adversarial action, such as exploiting one public-facing service or dropping one payload, matching the granularity at which ATT\&CK defines individual techniques. Each atomic behavior is a single brief sentence, phrased in the report's own verbs and technical terms, with technical strings such as domains and file paths kept unchanged.

Second, to ground every atomic behavior in report evidence, BEACON uses a deterministic locator without relying on the LLM to identify evidence locations. Locating evidence requires the output to match the report text exactly, while LLM outputs are generated rather than retrieved from the text, so generated evidence is not guaranteed to exist in the report. A deterministic locator instead guarantees that every evidence span is taken from the report, avoiding this form of hallucination. The locator scores every bounded window of consecutive sentences by the tokens shared with $b$, weighting each match inversely to its frequency in the report so that rare terms dominate. Before scoring, the locator stems the behavior and every report sentence, and removes stopwords and the report's most frequent tokens, which carry little behavior-specific information. Matched two- and three-word phrases receive an additional bonus, since phrase matches are less likely to be coincidental than single-token matches. The highest-scoring window becomes the evidence span $s(b)$.

Third, since ATT\&CK contains hundreds of techniques with fine-grained definitional differences and a single candidate can easily miss the correct one, the LLM reads $b$ together with $s(b)$ and proposes multiple candidate ATT\&CK techniques. The proposal is responsible for recall, so the candidate techniques only need to be relevant to the evidence. Candidates that do not exist in the ATT\&CK version BEACON uses are removed to suppress hallucination. For example, the LLM can propose a revoked ATT\&CK technique that its training data still contains.

Fourth, BEACON verifies the remaining candidates against their official definitions in the MITRE ATT\&CK catalog, through an LLM-as-a-judge mechanism~\citep{zheng2023judging}. The LLM receives the atomic behavior, its evidence span, and each candidate's official ATT\&CK name and definition, and judges each candidate separately. To keep the scores consistent across judgments, the LLM assigns confidence at predefined levels. Higher levels require the evidence to satisfy more of the technique's definition. To keep the judgment within the evidence and prevent weak evidence from passing, we instruct the LLM not to infer from its own knowledge and require conservative scoring when the evidence is ambiguous. Candidates whose confidence exceeds a threshold become anchors, each supported by evidence within the report and carrying the official ATT\&CK ID of its technique, such as T1190, which is shared across all reports.

\subsubsection{Contextual Entity and IoC Extraction}
Since the LLM can miss entities when it reads a long report in one pass, BEACON splits the report into overlapping windows of consecutive sentences, so that entities at window boundaries have enough context to be extracted. The LLM then extracts contextual entity and IoC candidates with types from each window, according to the predefined ontology. Unlike an atomic behavior, which is rephrased by the LLM, a contextual entity or IoC is an exact substring of the report. BEACON therefore lets the LLM specify the evidence sentences of each candidate directly, since an incorrect specification can be easily checked by string matching. Since the candidates are LLM outputs and can differ from the report text in characters such as letter case, BEACON applies word-level matching to candidates that fail the exact check, which normalizes the candidate and each sentence and accepts a sentence if its word-set Jaccard overlap with the candidate exceeds a threshold. To prevent an incorrect location from removing a real entity, BEACON searches the rest of the window in the same way if all named sentences fail both checks. Candidates found nowhere in the report are removed as a form of hallucination.

To reduce duplicate verification, contextual entities and IoCs are normalized and deduplicated first. Since IoC spellings are stable but often obfuscated by type-specific conventions, a per-type deterministic normalizer converts each IoC into a normalized form, such as removing the brackets in a domain written as \texttt{example[.]com}. IoCs with identical normalized forms then merge directly. Contextual entities have no external standard form, so BEACON only lowercases names, removes extra spaces, and merges exact type–name duplicates, leaving the remaining cases to the consolidation stage. After deduplicating, BEACON verifies contextual entities and IoC candidates. Since the same contextual entity can play different roles in different contexts, the LLM verifies each contextual candidate separately against its own evidence, confirming that it matches its type definition in the ontology, that it is a specific entity rather than a generic description like ``a popular website'', and that it is neither marketing language, nor defensive advice, nor report metadata such as authors and publication dates. When uncertain, the LLM rejects the candidate. IoC candidates instead have standardized forms after normalization, so the LLM reviews them in one pass against the full report, reporting only values to refuse or types to correct.

\subsubsection{Anchored Attachment}
Since there is no explicit relation between entities and ATT\&CK techniques to extract like the triplets, BEACON adds all edges after nodes. Since verifying every entity–technique pair with the LLM is costly, and an entity that participates in a behavior tends to be described near that behavior in the report, BEACON proposes a candidate edge wherever the entity's and the anchor's evidence spans share at least one sentence. The LLM receives both spans and the anchor's ATT\&CK definition, and judges whether the entity participates in that behavior. A candidate is accepted only when the shared sentences describe the relation explicitly or support it directly. It is rejected when the entity and the behavior merely share a sentence or the report topic, so that co-occurrence alone does not produce an edge.

A report often names its central entity, a threat actor, malware family, or campaign, without repeating it in every behavior description. The evidence spans of these behaviors therefore often do not contain the entity, and the edges between the central entity and its behaviors are missed. BEACON therefore treats ThreatActor, Campaign, and MalwareFamily as report-central types and processes entities of these types in two passes. The LLM first judges, based on the report text and the ATT\&CK techniques with their definitions, whether the entity is the central subject of the report, that is, whether the report describes the behaviors as performed by this entity. Entities that pass this check are paired with every technique anchor, each pairing verified by the same process.

\subsection{Consolidation Stage}
\label{sec:consolidation}

To merge the report-level graphs $\{G_1, \dots, G_n\}$ into $G^{*}$, we design a hierarchical alignment strategy that applies alignment signals in decreasing order of determinism, and each merge proposed by a later signal runs on the graph already merged by earlier signals. The LLM thus judges a smaller set of ambiguous names, reducing both cost and the chance of misclassification and hallucination. Nodes merged by earlier signals also combine the evidence of their members, giving the LLM richer context when it verifies the candidates from less deterministic signals.

\paragraph{Deterministic Matching}
Since the anchoring stage has given each technique anchor its official ATT\&CK ID, two anchors merge when they carry the same ID. Two IoCs merge when they share a type and their values are identical after per-type normalization. For a few IoC types, identical strings do not reliably indicate the same real-world entity, so BEACON keeps IoCs of these types unmerged. For example, a mutex name, the name a program assigns to a lock, can be reused by unrelated programs. Contextual entities with the same type and normalized name merge directly. For threat actor names in the same numbered naming scheme (a prefix such as APT, UNC, FIN~\citep{mandiantThreatActorClassifications} or TA~\citep{proofpoint2025craftycamel} followed by a number), BEACON applies an additional type-specific normalization such as removing case and leading-zero differences, and two names merge when both the prefix and the number match. Every merged node keeps the source report of each member (Figure~\ref{fig:overview}, right), so any merged node can be traced back to the sentences and sources that support it.

\paragraph{Similarity-based Matching}
Since multiple vendors independently report on the same threat under different naming conventions, the names of the same entity can remain different even after normalization, so contextual entities require more than deterministic rules. There are three more candidate proposal channels, and each channel uses one alignment signal to propose candidate pairs. The pairs from all channels are combined into per-type connected components as candidate clusters that the LLM later verifies.

The \emph{character} channel targets spelling variations such as ``Cl0p'' versus ``Clop''. Since entity types differ in the string patterns of their names, the channel uses four measures, each suited to a different pattern: Jaro–Winkler for short names that differ near the end, such as a trailing version digit; normalized Levenshtein for names that differ by a few characters, such as Cl0p and Clop; trigram cosine for multi-word names, since its score remains high when words are inserted or reordered; and trigram Jaccard, a stricter overlap measure that ignores repetition, for short names where repeated substrings inflate similarity.
Each entity type is scored by a weighted subset of the four measures, with its own normalization before scoring. For example, campaign names remove the ``Operation'' prefix and are scored by trigram cosine, Jaro–Winkler, and normalized Levenshtein. A pair becomes a candidate when its score exceeds a threshold. The channel uses a higher threshold when names are short, as short names are more likely to be similar by coincidence. Two additional rules handle CTI naming patterns where character similarity can be misleading. First, threat actor pairs in the same numbered naming scheme but with different numbers are excluded from scoring, since names like APT41 and APT40 are nearly identical in spelling but never the same entity. Second, a short all-letter name is paired with any multi-word name whose initials spell it, such as IRS and Internal Revenue Service, which captures acronym pairs that the four measures miss.

The \emph{embedding} channel handles entities whose names have character similarity too low for the character channel but may still be semantically related, such as ``Clop ransomware group'' and ``Cl0p'' for the same actor. A text-embedding model encodes each entity name, and a pair becomes a candidate when its embedding cosine similarity exceeds a threshold. To avoid comparing all pairs within a type, the channel computes similarity only for name pairs with basic lexical overlap: an identical normalized name, a shared word, or a trigram Jaccard or Jaro--Winkler score above a low threshold. Entities with numbered naming schemes are excluded from this channel, since different numbers always indicate different entities, a rule that embedding similarity can violate.

The \emph{technique-neighborhood} channel targets entities with unrelated names, which neither the character nor the embedding channel can propose, such as Graceful Spider and Cl0p for the same actor. For each entity $v$, the channel collects its ATT\&CK technique neighborhood $N_a(v)$ that $v$ attaches to. Since entities sharing multiple attack behaviors are likely the same entity, two entities become a candidate pair when their neighborhoods share at least two ATT\&CK techniques. This channel runs last, since earlier merges combine member neighborhoods and expose candidates that previously shared too few ATT\&CK techniques.

Every candidate cluster proposed by the three channels goes through LLM verification. The LLM receives each member's name, type, source report, and evidence spans, and gives confidence judgments at predefined levels on two granularities. Since a candidate cluster can contain members that do not belong while the rest are correct, a node-level judgment removes the members that do not fit. Since a candidate cluster is a connected component of pairwise candidates, it can contain names that were never proposed as a pair. That is, if A pairs with B and B with C, all three enter one cluster even when A and C are distinct. A cluster-level judgment therefore decides whether all members refer to one real-world entity.

\paragraph{Iterative Closure.}

A merge combines the ATT\&CK technique neighborhoods of its members, so entities that previously shared too few techniques can then become candidates through the technique-neighborhood channel. To align the entities that only become candidates after earlier merges, the channel iterates until no new merge passes verification. Since merging creates no new names, every pair the character and embedding channels could match was already available initially, so these channels do not iterate. Combining neighborhoods also increases the overlap between unrelated entities, but overlap only produces candidates, and every candidate passes the same LLM verification before it is accepted. The iteration always stops, since each round either merges nodes or, when no candidate passes verification, leaves the neighborhoods unchanged so that no new candidate can be proposed.

\section{Evaluation}

\subsection{Evaluation Setup}
\label{sec:setup}

\paragraph{Datasets}
We construct and release two datasets, \textbf{BEACON-Single} for report-level extraction and \textbf{BEACON-Group} for cross-source consolidation. Each instance is independently labeled by two experts, and disagreements are adjudicated by a senior annotator. 

BEACON-Single contains 150 reports from 15 publishers, with 4,966 annotated node instances and 3,429 edge instances. BEACON-Group organizes 100 reports from 31 publishers into 33 groups, each covering the same or closely related threats. The gold-standard consolidation merges the 2,908 report-level nodes into 1,950 consolidated nodes, of which 528 combine nodes from multiple reports, and the 2,278 edges into 1,537.

\paragraph{Research questions.}
Our evaluation answers three questions:
\begin{itemize}
    \item RQ1: does the space-anchoring stage accurately extract all three node types and their edges?
    \item RQ2: does the consolidation stage correctly align entities across sources?
    \item RQ3: how much do verification in the space-anchoring stage and each channel in the consolidation stage contribute?
\end{itemize}

\paragraph{Baselines.}
For RQ1 we compare five systems covering both lines of CTI extraction. From the line focusing on contextual information, CTINexus extracts ontology-constrained relation triplets end-to-end with an LLM. From the line focusing on behaviors, AttacKG+ uses an LLM to construct temporal behavior graphs and map them to ATT\&CK techniques; SoK-NER is a rule- and resource-based ATT\&CK matching pipeline; and SIGMERGE-TTP is a supervised BERT-based multi-subsequence classifier over a fixed ATT\&CK label set, whose recall upper bound on our dataset is 98.72\%. The step-guided LLM uses the same model, input, and target ontology as BEACON, but produces all outputs in one guided call without the anchoring pipeline. For RQ2 the baselines cover the three alignment signals without technique anchors. CTINexus clusters entities of the same type whose embedding similarity exceeds a threshold. Since its released implementation deduplicates within one report, we pool the report-level nodes of all reports in a group into one input, so that it deduplicates across them. ComEM uses an LLM to judge candidate pairs retrieved by lexical similarity and to select the best match among the remaining candidates. GA-MGM applies general-domain structural matching, jointly aligning all graphs of a group using fixed text similarities and graph topology. All baseline outputs are converted to our ontology by deterministic rules without access to the gold annotations.

\paragraph{Metrics.}
All results are reported as micro precision, recall, and F1. For RQ1, a predicted node or edge is correct when it matches a gold node or edge of the same type in the same report, under one-to-one matching. For RQ2, we score the final clusters in merge-link units: a cluster of size $k$ contributes $k-1$ links, and a predicted cluster $P$ that overlaps a gold cluster $G$ contributes $\max(|P \cap G|-1,\,0)$ correct links. This metric rewards partially recovered clusters, penalizes over-merging, and avoids the quadratic weight that pairwise counting gives to large clusters. Overall consolidation results aggregate all three node types. All consolidation systems take the same gold report-level nodes and edges of each group as input, so RQ2 measures alignment quality independently of extraction quality.

\paragraph{Implementation.}

Every system that requires an LLM uses GPT-4o~\citep{openai2024gpt4o}. All systems use the same ATT\&CK v17.1 snapshot, and all thresholds were tuned once on a small held-out development set and then fixed. Each reported number comes from a single run per configuration, and we quantify sampling uncertainty with 95\% percentile bootstrap confidence intervals (10{,}000 resamples over reports for RQ1 and over groups for RQ2) and paired bootstrap tests for the main comparisons.

\subsection{Results}

\subsubsection{RQ1: Anchoring-Stage Extraction}
\label{sec:rq1}

\begin{table*}[t]
  \centering
  \scriptsize
  \setlength{\aboverulesep}{0pt}
  \setlength{\belowrulesep}{0pt}
  \renewcommand{\arraystretch}{1.15}
  \caption{RQ1 space-anchoring extraction: micro precision, recall, and F1 (\%) for the three node types and two edge types. Best results in bold, second-best underlined. ``--'' marks outputs that a baseline does not natively produce.}
  \label{tab:rq1-anchoring}
  \resizebox{\textwidth}{!}{%
  \begin{tabular}{l|ccc|ccc|ccc|ccc|ccc}
    \toprule
    \rowcolor{headteal}
    & \multicolumn{3}{c|}{\textbf{Contextual entity}}
      & \multicolumn{3}{c|}{\textbf{TTP}}
      & \multicolumn{3}{c|}{\textbf{IoC}}
      & \multicolumn{3}{c|}{\textbf{Ctx--TTP}}
      & \multicolumn{3}{c}{\textbf{IoC--TTP}} \\
    \rowcolor{headteal}
    \textbf{Method} & P & R & F1 & P & R & F1 & P & R & F1 & P & R & F1 & P & R & F1 \\
    \midrule
    \rowcolor{rowgray}
    CTINexus~\citep{cheng2025ctinexus}
      & 40.93 & 31.26 & 35.45
      & -- & -- & --
      & 41.81 & 14.04 & 21.02
      & -- & -- & --
      & -- & -- & -- \\
    AttacKG+~\citep{zhang2025attackg+}
      & 22.85 & 24.82 & 23.80
      & 54.70 & 44.99 & \underline{49.37}
      & 9.99 & 24.48 & 14.19
      & -- & -- & --
      & -- & -- & -- \\
    \rowcolor{rowgray}
    SoK-NER~\citep{buchel2025sok}
      & -- & -- & --
      & 26.61 & 31.34 & 28.78
      & -- & -- & --
      & -- & -- & --
      & -- & -- & -- \\
    SIGMERGE-TTP~\citep{caitexts}
      & -- & -- & --
      & 42.06 & \underline{52.72} & 46.79
      & -- & -- & --
      & -- & -- & --
      & -- & -- & -- \\
    \rowcolor{rowgray}
    Step-guided LLM
      & \underline{73.23} & \underline{34.64} & \underline{47.03}
      & \underline{68.22} & 27.57 & 39.27
      & \textbf{88.65} & \underline{42.98} & \underline{57.89}
      & \underline{39.91} & \underline{11.07} & \underline{17.33}
      & \underline{38.69} & \underline{17.09} & \underline{23.70} \\
    \midrule
    \rowcolor{white}
    BEACON (Ours)
      & \textbf{90.66} & \textbf{68.08} & \textbf{77.77}
      & \textbf{72.51} & \textbf{84.60} & \textbf{78.09}
      & \underline{83.68} & \textbf{79.32} & \textbf{81.44}
      & \textbf{59.10} & \textbf{62.64} & \textbf{60.82}
      & \textbf{56.68} & \textbf{45.56} & \textbf{50.52} \\
    \bottomrule
  \end{tabular}}
\end{table*}

Table~\ref{tab:rq1-anchoring} reports the space-anchoring results. BEACON achieves the best F1 on all five output types (78.7\% overall node F1, 95\% CI [77.4, 79.9]) and exceeds every baseline on each type by at least 23\%. The step-guided LLM, the one baseline that covers all five outputs, shares BEACON's model, so this advantage comes from the pipeline of the space-anchoring stage rather than from the model.

\paragraph{Technique anchors.}
For BEACON, TTP extraction measures the quality of the technique anchors that the consolidation stage builds on. SIGMERGE-TTP has high recall but much lower precision, because the classifier, trained on official ATT\&CK text, assigns too many techniques to the broader behavior descriptions in real reports. AttacKG+ achieves the best F1 among the baselines but is limited by its multi-step pipeline, where triplet rewriting and tactic--technique labeling each lose information and accumulate errors. The step-guided LLM has the lowest recall, as it completes all extraction in one long guided call and misses many behaviors. BEACON's advantage is concentrated in recall (84.60), since decomposing the report into atomic behaviors allows multiple candidate techniques to be proposed for each behavior separately, while verification accepts a candidate only when its evidence span supports the technique's ATT\&CK definition.

\paragraph{Contextual entities and IoCs.}
CTINexus and AttacKG+ extract entities only when they appear in relation triplets, so entities without an explicit relation are not extracted. BEACON instead attaches entities to technique anchors without requiring entity--entity relations, which explains its higher recall on both contextual entities and IoCs. The difference is largest for CTINexus, whose IoC recall is 14.04, since IoCs usually appear in lists rather than in sentences describing relations, so a triplet extractor rarely finds a relation that includes them. AttacKG+ extracts only attack-relevant triplets, which further limits the range of contextual entities it covers, and part of its low IoC score comes from IoC types that partially overlap with ours. The step-guided LLM misses entities for the same reason it misses behaviors. On IoCs, the step-guided LLM has higher precision but extracts conservatively and misses over half of the gold IoCs, while BEACON's F1 comes from deterministic normalization and filtering.

\paragraph{Anchored edges.}
Edges are the hardest output, as a correct edge requires two correct endpoints and verified evidence for their relation. The step-guided LLM is the only baseline that produces edges of these types, since SoK-NER and SIGMERGE-TTP extract techniques alone, and CTINexus and AttacKG+ do not relate entities to techniques. IoC--TTP edges are the harder of the two, because IoCs in lists rarely share a sentence with a behavior's evidence span, so fewer candidate edges are proposed.

\subsubsection{RQ2: Cross-Source Consolidation}
\label{sec:rq2}

\begin{table}[t]
  \centering
  \scriptsize
  \setlength{\aboverulesep}{0pt}
  \setlength{\belowrulesep}{0pt}
  \renewcommand{\arraystretch}{1.15}
  \caption{RQ2 consolidation-stage micro precision, recall, and F1 (\%) for contextual entities and IoCs. Best results in bold, second-best underlined.}
  \label{tab:rq2-consolidation}
  \resizebox{\columnwidth}{!}{%
  \begin{tabular}{l|ccc|ccc}
    \toprule
    \rowcolor{headteal}
    & \multicolumn{3}{c|}{\textbf{Contextual entity}} & \multicolumn{3}{c}{\textbf{IoC}} \\
    \rowcolor{headteal}
    \textbf{Method} & P & R & F1 & P & R & F1 \\
    \midrule
    \rowcolor{rowgray}
    GA-MGM~\citep{wang2020graduated}
      & 31.42 & 45.51 & 37.17
      & 10.06 & 56.67 & 17.09 \\
    ComEM~\citep{wang2025match}
      & 62.01 & 68.27 & 64.99
      & 26.26 & 86.67 & 40.31 \\
    \rowcolor{rowgray}
    CTINexus~\citep{cheng2025ctinexus}
      & \underline{85.17} & \underline{82.85} & \underline{84.00}
      & \underline{32.61} & \textbf{100.00} & \underline{49.18} \\
    \midrule
    \rowcolor{white}
    BEACON (Ours)
      & \textbf{98.59} & \textbf{89.74} & \textbf{93.96}
      & \textbf{100.00} & \underline{96.67} & \textbf{98.31} \\
    \bottomrule
  \end{tabular}}
\end{table}

\paragraph{Full consolidation.}
Table~\ref{tab:rq2-consolidation} reports the consolidation results. BEACON achieves the best F1 on both node types (93.96 and 98.31) and exceeds every baseline by at least 9\%. Technique anchors, which account for 304 of the 958 gold merge-link units, merge deterministically by shared technique IDs and are therefore excluded from the comparison. CTINexus aligns entities by embedding similarity, which incorrectly merges IoCs with similar strings (32.61 precision, compared with BEACON's 100 under deterministic matching) and cannot propose aliases with unrelated names. ComEM cannot propose such candidates either, and the LLM cannot judge a pair that is not proposed, which limits recall. Its precision is low because both its BM25 blocking and its Flan-T5 match ranking prefer candidates with overlapping context, and its selector rarely chooses the no-match option, so distinct entities from one campaign are merged. Its IoC alignment applies no CTI-specific normalization and relies on the model's own knowledge of IoC formats, which results in a low F1 (40.31). GA-MGM aligns graphs by text similarity and graph topology, but it expects dense graphs where entities connect directly to each other, while the report graphs in this task are sparse, with edges only between entities and technique anchors, so it loses both precision and recall. Overall, shared technique neighborhoods are the one structural signal that holds across sources, and give BEACON its additional, domain-specific alignment advantage.

\begin{table}[t]
  \centering
  \scriptsize
  \setlength{\aboverulesep}{0pt}
  \setlength{\belowrulesep}{0pt}
  \renewcommand{\arraystretch}{1.15}
  \caption{RQ2 residual alignment: micro precision, recall, and F1 (\%) on the 220 gold contextual-entity alignment units that deterministic rules cannot align. Best results in bold, second-best underlined.}
  \label{tab:nondeterministic-alignment}  
  \resizebox{0.6\columnwidth}{!}{%
  \begin{tabular}{l|ccc}
    \toprule
    \rowcolor{headteal}
    \textbf{Method} & \textbf{P} & \textbf{R} & \textbf{F1} \\
    \midrule
    \rowcolor{rowgray}
    ComEM~\citep{wang2025match} & 39.66 & \textbf{75.00} & 51.89 \\
    GA-MGM~\citep{wang2020graduated} & 10.16 & 30.91 & 15.30 \\
    \rowcolor{rowgray}
    CTINexus~\citep{cheng2025ctinexus} & \underline{55.88} & 51.82 & \underline{53.77} \\
    \midrule
    \rowcolor{white}
    BEACON (Ours) & \textbf{95.12} & \underline{70.91} & \textbf{81.25} \\
    \bottomrule
  \end{tabular}}
\end{table}

\paragraph{Residual alignment.}
Table~\ref{tab:nondeterministic-alignment} evaluates the 220 gold contextual-entity alignment units that deterministic rules cannot align, entities whose names differ in spelling or are entirely unrelated across sources. A unit is kept when its members share neither the same normalized name nor the same type--name pair, so the filter removes only units whose names already match, which every compared system aligns correctly. BEACON retains 81.25 F1 on this subset (95\% CI $[72.9, 87.3]$), while CTINexus drops to 53.77 ($[45.5, 61.8]$), as the remaining cases concentrate on the aliases that differ most across sources. The paired difference is $+27.5$ F1 ($p < 10^{-4}$). Embedding similarity still aligns roughly half of these units, but it cannot align names that are semantically unrelated. ComEM achieves 51.89 F1, with the highest recall in the table (75.00) but low precision (39.66), since its selector merges nearly every proposed candidate. GA-MGM drops to 15.30, as both of its signals fail on this subset. Its text similarities cannot match unrelated names, while names in the same numbered naming scheme look similar and are matched, and its graph topology does not help on these sparse report graphs. On this subset, removing the technique-neighborhood channel drops F1 from 81.25 to 64.85, since it proposes the candidates that both character and embedding similarity miss.

\begin{figure}[t]
  \centering
  \includegraphics[width=\columnwidth]{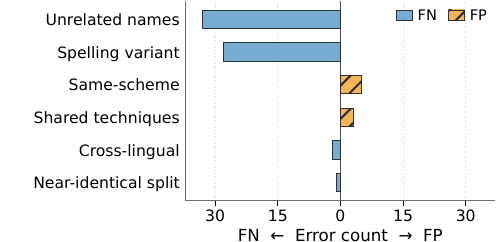}
  \caption{Contextual-entity error analysis under the merge-link metric. Missed merges (FN) are shown on the left, and false merges (FP) on the right.}
  \label{fig:rq2-error-anatomy}
  \Description{A horizontal diverging bar chart showing contextual-entity alignment errors by category. False negatives are shown as blue bars extending left from zero, and false positives are shown as orange hatched bars extending right. Unrelated names and spelling variants account for the largest numbers of false negatives. Same-scheme and shared techniques are the main sources of false positives. Cross-lingual cases and near-identical entity splits contribute only a small number of false negatives.}
\end{figure}

\paragraph{Error analysis.}
Figure~\ref{fig:rq2-error-anatomy} shows the contextual-entity errors under the merge-link metric, where missed merges are about ten times more common than false merges. Over half of the missed alignments come from entities whose names are entirely unrelated, such as SPIKEDWINE and APT29, and whose technique neighborhoods share too few techniques, so none of the three channels can propose the pair. Nearly half of the remaining missed alignments involve abbreviations and spelling variants, but their main cause is the LLM verification, which is conservative and rejects candidates with weak supporting evidence, as with the pair IRS and Internal Revenue Service, proposed by the character channel but rejected. On the false-merge side, most cases merge similar names or names in the same numbered naming scheme, such as Storm-1175 with Storm-0506. The exclusion rule covers a fixed list of prefixes that does not include Storm, showing that the naming schemes in CTI are too diverse to enumerate. The remaining false merges combine two actors that share techniques but are different entities, as with Black Basta and Cactus.

\subsubsection{RQ3: Component Attribution}
\label{sec:ablation}

\begin{table}[t]
  \centering
  \scriptsize
  \setlength{\aboverulesep}{0pt}
  \setlength{\belowrulesep}{0pt}
  \renewcommand{\arraystretch}{1.15}
  \caption{RQ3 space-anchoring verification ablation. Each row removes one verification component and reports the change in micro P/R/F1 (\%) on the output type that the component applies to, relative to full BEACON (Table~\ref{tab:rq1-anchoring}).}
  \label{tab:rq1-ablation}
  \resizebox{\columnwidth}{!}{%
  \begin{tabular}{l|l|ccc}
    \toprule
    \rowcolor{headteal}
    \textbf{Removed component} & \textbf{Scope} & \textbf{$\Delta$P} & \textbf{$\Delta$R} & \textbf{$\Delta$F1} \\
    \midrule
    \rowcolor{rowgray}
    w/o Entity review     & Ctx                  & $-14.06$ & $+7.67$  & $-1.59$  \\
    w/o TTP verification  & TTP                  & $-42.78$ & $+8.54$  & $-33.02$ \\
    \rowcolor{rowgray}
    w/o IoC filtering     & IoC                  & $-24.75$ & $+11.76$ & $-9.88$  \\
    w/o Edge confirmation & Ctx--TTP             & $-15.42$ & $+2.71$  & $-8.46$  \\
    \rowcolor{rowgray}
    w/o Edge confirmation & IoC--TTP             & $-14.02$ & $+4.11$  & $-4.62$  \\
    \bottomrule
  \end{tabular}}
\end{table}

\paragraph{Anchoring verification.}
Table~\ref{tab:rq1-ablation} reports the ablation of each verification component in the space-anchoring stage. Every removal raises recall but lowers precision more, so verification suppresses misclassification and hallucination at a smaller cost in recall. TTP verification has the largest effect, confirming the value of verifying candidate techniques against official ATT\&CK definitions. Entity review has the most balanced effect, leaving F1 nearly unchanged. IoC filtering changes both precision and recall substantially, as IoCs such as paths and IPs are easy to extract by format but not all of them are attack-related, so the filter removes many unrelated ones while occasionally removing valid ones. Removing edge confirmation lowers precision substantially on both edge types, which shows that many candidate edges proposed from shared sentences reflect co-occurrence rather than participation.

\begin{table}[t]
  \centering
  \scriptsize
  \setlength{\aboverulesep}{0pt}
  \setlength{\belowrulesep}{0pt}
  \renewcommand{\arraystretch}{1.15}
  \caption{RQ3 consolidation-stage ablation using micro metrics (\%). Each row removes one candidate proposal component. Best results in bold, second-best underlined.}
  \label{tab:rq2-ablation}
  \resizebox{0.85\columnwidth}{!}{%
  \begin{tabular}{l|ccc|c}
    \toprule
    \rowcolor{headteal}
    & \multicolumn{3}{c|}{\textbf{Contextual entity}} & \textbf{Overall} \\
    \rowcolor{headteal}
    \textbf{Configuration} & P & R & F1 & F1 \\
    \midrule
    \rowcolor{rowgray}
    w/o char measures
      & 98.39 & \underline{88.30} & 93.07 & 95.52 \\
    w/o char rules
      & 98.74 & 88.14 & \underline{93.14} & \underline{95.56} \\
    \rowcolor{rowgray}
    w/o embedding
      & \underline{99.03} & 81.57 & 89.46 & 93.30 \\
    w/o neighborhood
      & \textbf{99.42} & 81.89 & 89.81 & 93.52 \\
    \midrule
    \rowcolor{white}
    Full BEACON
      & 98.59 & \textbf{89.74} & \textbf{93.96} & \textbf{96.07} \\
    \bottomrule
  \end{tabular}}
\end{table}

\paragraph{Consolidation channels.}
Table~\ref{tab:rq2-ablation} reports the ablation of each proposal component in the consolidation stage. The full configuration achieves the best contextual and overall F1, and every removal lowers recall while leaving precision nearly unchanged, since the channels only propose candidates and the same LLM verification decides every merge. The embedding and neighborhood channels contribute the most to recall, by similar amounts. The neighborhood channel aligns the entities with unrelated names, which the embedding channel cannot propose. The two character-level components contribute less, as their easy cases are already resolved by deterministic normalization and their harder cases partially by the embedding channel. Precision is highest when the neighborhood channel is removed (99.42), indicating that the neighborhood channel proposes the least reliable candidates of the three.

\paragraph{Model sensitivity and cost.}
In anchoring, GPT-4o, GPT-4o-mini, and GPT-5.6-luna reach node F1 of 78.7, 52.7, and 76.7 at \$0.32, \$0.02, and \$0.31 and in 135, 194, and 271 s per report. In consolidation, they reach 96.1, 89.1, and 96.1 at \$0.020, \$0.001, and \$0.012 and in 11.5, 15.3, and 13.3 s per group. A smaller model trades accuracy for cost, while a comparable model reaches similar accuracy.

\section{Limitations}

BEACON depends on ATT\&CK coverage. A behavior that ATT\&CK does not define produces no anchor, so entities without anchors cannot be aligned by technique neighborhoods. Consolidation is evaluated on gold report-level graphs within groups of related reports, which does not reflect performance in end-to-end and open-world settings. BEACON makes more LLM calls than the baselines, and the comparison does not control for this budget.

\section{Conclusion}

We identified that existing approaches to constructing knowledge graphs from CTI reports extract only partial information and leave the cross-source setting unexplored. We proposed BEACON based on the insight that attack behaviors can be reliably mapped to MITRE ATT\&CK, while contextual entities and IoCs describe their participants and traces. The space-anchoring stage attaches contextual entities and IoCs to ATT\&CK technique anchors, placing per-report graphs in one canonical space. The consolidation stage merges them by signals applied in decreasing order of determinism, where the shared technique neighborhoods are critical for aligning entities with unrelated names. To suppress misclassification and hallucination, both stages follow a propose-then-verify paradigm grounded in report evidence. On two human-annotated datasets that we construct and release, BEACON outperforms all baselines.

\begin{acks}
This work is supported by the \grantsponsor{NSF}{National Science Foundation}{https://doi.org/10.13039/100000001} under grant~\grantnum{NSF}{2442171} and the Amazon Research Award (ARA). Any opinions, findings, and conclusions made in this paper are those of the authors and do not necessarily reflect the views of the \mbox{funding agencies.}
\end{acks}

\section*{Ethical Considerations}

BEACON supports defensive threat analysis by organizing publicly available CTI reports and grounding behaviors in public MITRE ATT\&CK definitions. It generates no attack procedures, exploit code, or operational instructions beyond what the source reports already disclose, and therefore introduces limited dual-use risk beyond the original public sources. The datasets are built from publicly released reports and contain no private telemetry or undisclosed victim data. Victim-related information is retained only where the source report publicly discloses it and the schema requires it. Annotators were informed of the research use of their work and compensated. Released materials preserve source attribution and comply with the original publishers' access terms, with redistribution-restricted text provided as metadata and derived annotations.

\bibliographystyle{ACM-Reference-Format}
\bibliography{reference/reference}

\end{document}